\documentclass[preprint,onecolumn,12pt]{article}
\usepackage[hmargin=3cm,vmargin=3.5cm]{geometry}
\usepackage{graphicx}
\usepackage{amsmath}
\usepackage{amssymb}
\usepackage{enumerate}
\usepackage{cite}
\usepackage{dcolumn}
\usepackage{bm}
\usepackage[colorlinks=true]{hyperref}
\usepackage{tikz}
\usetikzlibrary{calc,decorations.markings}

\usepackage{subfig}
\usepackage{amsfonts,amsthm,mathrsfs}
\usepackage{arydshln}

\newcommand{\bea}{\begin{eqnarray}}
\newcommand{\eea}{\end{eqnarray}}

\begin{document}

 \usetikzlibrary{decorations.pathmorphing, patterns,shapes}

 \title{Detectors interacting through quantum Fields: non-Markovian effects, non-perturbative generation of correlations and apparent non-causality }
 \author {Theodora Kolioni \footnote{kolioni@upatras.gr}  \; and  \; Charis Anastopoulos \footnote{anastop@physics.upatras.gr}\\
 {\small Department of Physics, University of Patras, 26500 Greece} }

\maketitle

\begin{abstract}
 We study the system of two localized detectors (oscillators) interacting through a massless quantum field in a vacuum state via an Unruh-DeWitt coupling. This system admits an exact solution is providing a good model for addressing fundamental issues in particle-field interactions, causality, and locality in quantum field measurements that are relevant to proposed quantum experiments in space. Our analysis of the exact solution leads to the following results. (i) Common approximations used in the study of analogous open quantum systems fail when the distance between the detectors becomes of the order of the relaxation time. In particular, the creation of correlations between remote detectors is not well described by ordinary perturbation theory and the Markov approximation. (ii) There is a unique asymptotic state that is correlated; it is not entangled unless the detector separation is of the order of magnitude of the wavelength of the exchanged quanta. (iii) The evolution of seemingly localized observables is non-causal. The latter is a manifestation of Fermi's two-atom problem, albeit in an exactly solvable system. We argue that the problem of causality requires a re-examination of the notion of entanglement in relativistic systems, in particular, the physical relevance of its extraction from the quantum vacuum.
\end{abstract}

\section{Introduction}
Understanding how spatially separated quantum systems interact via relativistic quantum fields  becomes increasingly important. Many proposed quantum experiments in space lie in the regime where relativistic effects are  important and may even provide tests of new physics \cite{rideout}. Our ability to construct entangled states of atoms at large separations will reach a regime where the  retarded propagation of photons will be a significant factor, thus, allowing us to explore experimentally the relations between entanglement and relativistic causality. Furthermore, the interplay between localization and causality is a source of long-standing puzzles  in the foundations of Quantum Field Theory (QFT).

In this paper, we study an {\em exactly solvable} model that allows us to address issues such as the above. The model consists of two harmonic oscillators interacting with a quantum field through the Unruh-DeWitt coupling \cite{Unruh76, Dewitt, HLL12}. The field lies initially at the vacuum state. The harmonic oscillators can be viewed as particle detectors or as crude approximations to atoms ($N$-level systems). We find and analyse the exact solution to the system, to conclude the following.
\begin{enumerate}[(i)]

\item Common approximations that are employed in the treatment of analogous  quantum systems (Markov approximation, Wigner-Weisskopf approximation, perturbative master equation) fail if the separation of the two detectors becomes of the order of relaxation time. In particular, the above approximations break down completely in processes that involve the exchange of information between far separated detectors. While this result is derived in a specific model system, its context its quite generic for open quantum systems. In particular, it suggests that at least some entangled states for atoms at large separations decay non-exponentially.

    \item There is a unique asymptotic state of the system. This state is correlated, however, correlations are suppressed at large separations between the two detectors. For distances of the order of the wavelength of the exchanged quantum, the asymptotic state is entangled. The generated entanglement evolves significantly at times of the order of the relaxation scale. Hence, this model leads to different predictions about entanglement generation from treatments  that rely on time-dependent perturbation theory and do not incorporate back-reaction.

    \item If we assume that the variables pertaining to detectors are localized quantum observables, then the reduced dynamics of the detector are non-causal. This is a manifestation of the famous Fermi two-atom problem---see, below. Having an exact solution allows us to show that this behaviour is not an artefact of an approximation in the derivation of the dynamics. Rather, its origins are kinematical: we need to identify new observables that also involve the field degrees of freedom in order to describe localized measurements. This conclusion implies that  entanglement generated between the detectors may  not be a physically meaningful quantum resource to harvest.

\end{enumerate}
The  context of our results is the following.

\medskip

\noindent {\em Non-Markovian dynamics.}
A localized quantum system, such as an atom, in an excited state decays to the vacuum through its interaction with a quantum field, even if the latter is in the vacuum state. Such decays are typically exponential. When the system is treated in the theory of open quantum systems \cite{Davies, BrePe07}, the exponential decay law arises as a consequence of Markovian open system dynamics.

Markovian dynamics are generic for weak coupling of the system to environment. The second-order Markovian master equation becomes exact at the van Hove limit \cite{vanHo}. In this limit, the system-environment coupling $\lambda$ goes to zero,  while the rescaled time $\lambda^2 t$ remains constant. This limit provides an excellent approximation for a large class of systems, especially in atom optics. However, comparison with exactly solvable models---as, for example, in quantum Brownian motion \cite{HPZ}---shows many regimes in which the second order master equation fails.  In particular, the van Hove limit may not be physically relevant when  the open system dynamics are characterized by several  long-time scales. This occurs, for example,  if the environment has  resonance frequencies or thresholds \cite{Ana19}. In this paper, we present another case of failure of the Markov approximation, when the time-scale of retarded propagation is of the same order of magnitude with the decay / dissipation time.

The study of non-Markovian dynamics in open quantum systems has seen increased emphasis in recent years, because of the relevance of non-Markovian behavior to many different physical contexts, for example, condensed matter physics, quantum control, quantum biology and quantum optics---see,  Ref. \cite{VA17} and references therein.  Our ability to prepare entangled states in multipartite systems provides novel technical and conceptual challenges to the theory of open quantum systems, because they go beyond the traditional paradigm of a central, localized system weakly interacting with an environment.

Consider, for example, two atoms prepared in an entangled state,  separated by distance $r$ and interacting with a quantum electromagnetic field.
For small separations, this system is well described by the second order master equation---see, for example, Ref. \cite{FiTa}. However, as the separation increases, approximations involved in the derivation of the master equation break down, for example, the Rotating Wave Approximation \cite{Aga71, FCAH}. When  $r$ becomes comparable to the decay time  $\Gamma^{-1}$, the van Hove limit stops being  a useful approximation, because it misrepresents strong effects due to retarded propagation. Simply by analysing the mathematical assumptions involved in the Markov approximation, we expect the decay of an entangled pair of atoms to be strongly non-Markovian  when $\Gamma r$ becomes of order unity of larger.  This expectation is verified by our analysis.

Note that this breakdown of Markovian behaviour is a non-perturbative effect: $\Gamma$ is proportional to the coupling constant squared, but we can always find a distance $r$ such that $\Gamma r \sim 1$. For atomic states relevant to entanglement experiments,  the relevant length scale may be of the order of hundreds of meters or kilometers. Hence, the breakdown of Markovianity appears at scales relevant to macroscopic quantum phenomena.

 \medskip

 \noindent {\em Fermi's two-atom system.} The two-atom system is a classic model for propagation of information through quantum fields. It was first studied by
  Fermi \cite{Fermi}.
Fermi assumed that at time $t = 0$,  atom A is in an excited state and atom B in the ground state. He asked when B will notice A and move from its ground state. In accordance with Einstein locality, he found that this happens only at time greater than $r$. It took about thirty years for Shirokov to point out that Fermi's result is an artefact of an approximation \cite{Shiro}.

Several studies followed with conclusions depending on the approximations used \cite{Fermiproblem}. It was believed that non-causality is due to the use of bare initial states, and that it would not be present in a renormalised theory. However, Hegerfeldt showed that non-causality is generic \cite{Heg,Heg2}, as it depends only on the assumption of energy positivity and on the existence of systems that are localized in disjoint spacetime regions---see, also the critique in \cite{Buch}. The two-atom problem is a genuine problem of quantum theory that pertains to the definability of local observables and the meaning of locality in relation to quantum measurements.

\medskip

\noindent{\em Entanglement generation.} It is well known that two systems that do not directly interact may become entangled through their interaction with a third system. This general result also applies to localised systems (detectors) interacting with the quantum field. The detectors may develop entanglement even if the field lies on its ground state \cite{Rez13}. This process is called {\em entanglement harvesting} and it has been extensively studied for different initial detector states, detector trajectories, or spacetime geometries---see, for example,  \cite{S.M.M.15, K.M.15, M.S.T.}. Interestingly,  this process of entanglement creation may also take place between objects that remain spacelike separated, i.e., in some models, entanglement is seemingly generated outside the lightcone \cite{MaSp06, Fr08, LH10}.

However, it is far from obvious that the usual notion of entanglement, defined with reference to non-relativistic physics, is an appropriate quantum resource relativistic systems described by QFT. A proper quantum resource should be compatible with strong locality and causality constraints on acceptable physical observables that are required by QFT. Indeed, Fermi's problem is an indication that special care is needed in identifying acceptable local observables in a relativistic quantum system.
\medskip

\noindent {\em Our model.}
In this paper, we study the causal propagation of information between separated Unruh-DeWitt (UdW) detectors \cite{Unruh76, Dewitt, HLL12}, rather than between two atoms. An Unruh-DeWitt detector is a pointlike quantum system that interacts with  a quantum scalar field through a dipole coupling that mirrors the coupling of atoms to the electromagnetic field.

The main benefit of using the UdW detectors for studying information transfer in QFT is that they admit exact solutions.  In particular, if (i) the self-Hamiltonian of each detector corresponds to a harmonic oscillator, and (ii) the initial state of the field is Gaussian, then the system of $N$ detectors interacting with the quantum field is mathematically equivalent with a Quantum Brownian Motion (QBM) model   \cite{CaLe} for $N$ oscillators in a bath modeled by harmonic oscillators . This QBM model is exactly solvable \cite{HPZ, AKM10, FRH11}. Hence, we can compare the predictions of any  approximation with those of the exact solution. The model considered here has also been studied by Lin and Hu \cite{LH08}---see, also \cite{LCH08, LH10} for the same Hamiltonian but different  detector trajectories. Ref. \cite{LH08} employs a very different approximation scheme, and focuses on a different set of issues. Entanglement generation is a common issue, and there our results are compatible. However, we differ on the analysis of causality.

 \medskip

 The structure of this paper is the following. In Sec. 2, we present the general solution to the QBM model with $N$-system oscillators interacting with an environment, and the show that the system of two detectors interacting through a scalar field is a special solution. In Sec. 3, we find the explicit solution to the two-detector system and prove that the Markov approximation breaks down completely for the transfer of information between remote detectors. In Sec. 4, we identify a unique asymptotic state that is correlated, and show that it is entangled at small separations. In Sec. 5, we show that this model manifests the same non-causal behaviour with Fermi's two atom system, and we discuss the implications, and how causality can be restored. Sec. 6 concludes the paper.

\section{The model}

\subsection{ QBM in a multi-partite system}

\subsubsection{The Hamiltonian}
We consider a system of N harmonic oscillators of masses $M_{\alpha}$ and frequencies $\Omega_{\alpha}$ interacting with a heat bath. The bath is modelled by a set of harmonic oscillators of masses $m_{i}$ and frequencies $\omega_{i}$. The Hamiltonian of the total system is

\bea
\hat{H}=\hat{H}_{syst}+\hat{H}_{env}+\hat{H}_{int}
\eea
where
\bea
\hat{H}_{syst} &=&\sum_{\alpha}\left(\frac{1}{2M_{\alpha}}\hat{P}_{\alpha}^2+\frac{M_{\alpha}\Omega _{\alpha}^2}{2}\hat{X}_{\alpha}^2\right),\\
\hat{H}_{env}&=&\sum_{i}\left(\frac{1}{2m_{i}}\hat{p}_{i}^2+\frac{m_{i}\omega_{i}^2}{2}\hat{q}_{i}^2\right),\\
\hat{H}_{int}&=&\sum_{i}\sum_{\alpha}c_{i\alpha}\hat{X}_{\alpha}\hat{q}_{i},
\eea
where $c_{i\alpha}$ are coupling constants.

Since the total Hamiltonian is quadratic with respect to all positions and momenta, the evolution operator $e^{-i \hat{H}t}$ can be explicitly constructed, and its position matrix elements are Gaussian.

We consider a factorised initial condition $\hat{\rho}_{sys}\otimes \hat{\rho}_{env}$ for the total system. If $\hat{\rho}_{env}$ is Gaussian, then the reduced density matrix propagators can be computed explicitly. For $N = 1$, the reduced dynamics leads to the Hu-Paz-Zhang master equation \cite{HPZ}.

     In general, the assumption of a factorized initial condition between field and detectors is meaningful only as far as the field modes with energies of the order of the frequencies $\Omega_{\alpha} $ is concerned. There is no preparation that can enforce separability for photons at the infra-red and ultra-violet edges of the spectrum. However, a non-factorized initial condition does not allow us to consider general initial states for the field \cite{Pechukas} and in many model systems, including QBM,  the  effect of the non-factorizing initial state die out after a time-scale of the order of a high-frequency cut-off \cite{Romero}.

\subsubsection{The Wigner function propagator}

In this paper, we will employ the solution to the multi-partite QBM model in the Wigner representation \cite{HY, FOC, AKM10}---another form of the general solution is found in  \cite{FRH11}. The Wigner function for the reduced density matrix is defined as
\bea
W(\mathbf{X},\mathbf{P}) & = & {\frac{1}{(2\pi)^{N}}}{\int d \zeta e^{-\imath\ \mathbf{P}\cdot \zeta}{\hat \rho}\left(\mathbf{X}+\frac{1}{2}\zeta,\mathbf{X}-\frac{1}{2}\zeta\right)}.
\eea

 We use the coordinates $\xi^a = (X_1, X_2,\ldots, X_N, P_1, P_2, \ldots, P_N)$ on phase space; the Wigner function is expressed as $W(\xi)$. Dynamics in the Wigner picture is implemented by the Wigner function propagator   $ K_{t}(\xi_{f},\xi_{0} )$, namely, a kernel that evolves the initial Wigner function $W_0$ to the Wigner function $W_t$ at time $t$,
\bea
W_{t}(\xi_f)  =  \int \frac{ d ^{2N}\xi_{0}}{(2\pi)^{N}} K_{t}\left(\xi_{f},\xi_{0}\right) W_{0}(\xi_{0}).  \label{equa1}
\eea

For QBM models, the Wigner function propagator is Gaussian. The most general form of a Gaussian propagator is

\bea
K_{t}(\xi_{f},\xi_{0}) & = & \frac{\sqrt{detS^{-1}}}{\pi^{N}}\; exp\left[-\frac{1}{2}[\xi^{a}_{f}-\xi^{a}_{cl}(t)]S^{-1}_{ab}(t)[\xi^{b}_{f}-\xi^{b}_{cl}(t)]\right]. \label{propG}
\label{equa130}
\eea
where $S_{ab}$ is  positive definite matrix and
\bea
\xi^{a}_{cl}(t)=R^{a}_{b}(t)\xi^{b}_{0}.
\eea
The matrix $R^a_b$ defines the solution to the classical equations of motion. The matrix $S_{ab}$ determines the evolution of the environment-induced fluctuations. To see this, we consider the correlation matrix
\bea
V_{ab}: =\frac{1}{2}Tr[\hat{\rho}(\hat{\xi_{a}}\hat{\xi_{b}}+\hat{\xi}_{b}\hat{\xi}_{a})]-Tr(\hat{\rho}\hat{\xi_{a}})Tr(\hat{\rho}\hat{\xi}_{b}).
\label{equa11}
\eea
By Eq. (\ref{propG}),

\bea
V(t) =  R(t)V(0)R^{T}(t)+S(t) \label{equa2}
\eea
where $V_{0}$ is the correlation matrix of the initial state.

The explicit form of the matrices $R$ and $S$ was derived in Ref. \cite{AKM10}. They depend on two kernels, the {\em dissipation kernel},
 \bea
\gamma_{\alpha \alpha'}(s)=-\sum_{i}\frac{c_{i\alpha}c_{i\alpha'}}{2m_{i}\omega_{i}^{2}}sin(\omega_{i}s), \label{dissip}
\eea
and the {\em noise kernel},

\bea
\nu_{\alpha\alpha'}(s)=\sum_{i}\frac{c_{i\alpha}c_{i\alpha'}}{2m_{i}\omega_{i}^2}coth\left(\frac{\omega_{i}}{2T}\right)cos(\omega_{i}s), \label{noise}
\eea
that also characterize the path integral description of QBM \cite{CaLe, HPZ}.

The crucial step in the determination of the matrices $R$ and $S$ is to find the solution to the homogeneous part of the linear integrodifferential equation \cite{AKM10}

\bea
\ddot{\hat{X}}_{\alpha}(t)+\Omega_{r}^{2}\hat{X}_{\alpha}(t)+\frac{2}{M_{\alpha}}\sum_{\alpha'}\int _{0}^{t}ds \gamma_{\alpha\alpha'}(t-s)\hat{X}_{\alpha'}(s)= \sum_{i}\frac{c_{i\alpha}}{M_{\alpha}}\hat{q}_{i}^{0}(t). \label{sys}
\eea
 The solution of eq.(\ref{sys}) is
 \bea
 \hat{X}_{a}(t)&=&\sum_{\alpha}\left( \dot{u}_{\alpha \alpha'}(t)\hat{X}_{\alpha'}+\frac{1}{M_{\alpha'}}u_{\alpha\alpha'}(t)\hat{P}_{\alpha'} \right) \nonumber \\
 &+&\sum_{\alpha'}\frac{1}{M_{\alpha'}}\int_{0}^{t}ds u_{\alpha\alpha'}(t-s)\sum_{i}c_{i\alpha'}\hat{q}_{i}^{0}(s)
 \eea
where $u_{\alpha\alpha'}(t)$ is the solution of homogeneous part of eq.(\ref{sys}) with initial conditions $\dot{u}_{\alpha\alpha'}(0)=\delta_{\alpha\alpha'}$ and $u_{\alpha\alpha'}(0)=0$. Eq. (\ref{sys}) is essentially the classical equation of motion with a non-local-in-time dissipation term defined by the dissipation kernel.

Given the solution  $u(t)$, we define the matrix $R$ as
\bea
R=
\left(
    \begin{array}{c;{2pt/2pt}c}
        \dot{u}(t) & u(t)M^{-1} \\ \hdashline[2pt/2pt]
        M\ddot{u}(t) & M\dot{u}(t)M^{-1}
    \end{array}
\right)
\eea
 where $M= diag(M_1, . . . , M_N)$ is the mass matrix for the system.

The matrix elements of $S$ are given by
\bea
S_{X_{\alpha}X_{\alpha'}}&=&\sum_{\beta \beta'}\frac{1}{M_{\beta}M_{\beta'}}\int_{0}^{t}ds\int_{0}^{t}ds'u_{\alpha \beta}(s)\nu_{\beta\beta'}(s-s')u_{\beta'\alpha'}(s'), \label{sxx}\\
S_{P_{\alpha}P_{\alpha'}}&=&M_{\alpha}M_{\alpha'}\sum_{\beta\beta'}\frac{1}{M_{\beta}M_{\beta'}}\int_{0}^{t}ds\int_{0}^{t}ds'\dot{u}_{\alpha\beta}(s) \nu_{\beta\beta'}(s-s')\dot{u}_{\beta'\alpha'}(s'), \label{spp}\\
S_{X_{\alpha}P_{\alpha'}}&=&M_{\alpha'}\sum_{\beta\beta'}\frac{1}{M_{\beta}M_{\beta'}}\int_{0}^{t} ds \int_{0}^{t}ds'u_{\alpha\beta}(s)\nu_{\beta\beta'}(s-s')\dot{u}_{\beta'\alpha'}(s') \label{sxp}
\eea

  \subsection{Two UdW detectors}

We consider a system of two identical static harmonic oscillators of mass $M = 1$ and frequency $\Omega $ interacting with  a scalar field through the Unruh-DeWitt interaction Hamiltonian. The Hamiltonian of the total system  form, where we assume that the detectors are localized at $  {\pmb x} = {\pmb x}_1$ and ${\pmb x}   = {\pmb x}_2$

\bea
\hat{H}_{int}= \lambda \left( \int d^3 x\hat{\phi} ({\pmb x})\hat{q}_{1}\delta^3 ({\pmb x}-{\pmb x}_{1})+\int d^n x\hat{\phi} (x) \hat{q}_{2}\delta^3 ({\pmb x}- {\pmb x}_{2}) \right).\label{eq2}
\eea
where $\lambda$ is a coupling constant.

For a free scalar field, the total Hamiltonian
\bea
\hat{\phi}(x)=\int \frac{d^{3}k}{(2\pi)^{3}}\frac{1}{\sqrt{\omega_{k}}}(\hat{a}(k)e^{i{\pmb k} \cdot{\pmb x}}+\hat{a}^{\dagger}(k)e^{-i{\pmb k} \cdot{\pmb x}}),
\eea
 is a special case of the QBM Hamiltonian. The index $i$ corresponds to three momenta ${\pmb k}$, $m_i = 1$, $\omega_{\pmb k} = |{\pmb k}|$ and $c_{{\pmb k}_\alpha} = \frac{\lambda}{\sqrt{2\omega_{k}}}e^{\imath k x_{\alpha}}$.

It is straightforward to evaluate the dissipation kernel. By Eq. (\ref{dissip}),
\bea
\gamma(s) = \gamma_0(s) \left( \begin{array}{cc} 1&0  \\ 0&1\end{array} \right) + \gamma_r(s) \left( \begin{array}{cc} 0&1 \\ 1&0\end{array} \right)
\eea
where
\bea
\gamma_{0}(s)&=& -\frac{\lambda^{2}}{8\pi^{2}}\int_{0}^{\infty} dk sin(ks)\\
\gamma_{r}(s) &=&  -\frac{\lambda^{2}}{8\pi^{2}r}\left[\int_{0}^{\infty} dk \frac{sin(kr)\sin(ks)}{k}\right].
\eea
The function $\gamma_0(s)$ is the dissipation kernel of the one-detector system \cite{BaCa}. It must be regularized, for example, by introducing a high-frequency cut-off $\Lambda$. For $r \rightarrow 0$, $\gamma_r$ coincides with $\gamma_0$. In principle, we should introduce the same cut-off $\Lambda$ to $\gamma_r$, however $\gamma_r$ is little affected unless $r$ is of the order of $\Lambda^{-1}$ or smaller. Alternatively, we can regularize $\gamma_0$ be equating it with $\gamma_{r_0}$ for some $r_0 << r$.
By Eq. (\ref{noise}), the noise kernel is
\bea
\nu(s) = \nu_0(s) \left( \begin{array}{cc} 1&0 \\ 0&1\end{array} \right) + \nu_r(s) \left( \begin{array}{cc} 0&1 \\ 1&0\end{array} \right),
\eea
where
  \bea
  \nu_{0}(s) &=&   \frac{\lambda ^{2}}{8\pi }\delta(s)\\
  \nu_{r}(s) &=& \nu_{21}(s) = \frac{\lambda ^{2}}{32\pi r} \left[ \mbox{sgn}(r-s)+ \mbox{sgn}(r+s) \right].
\eea

\section{The classical equations of motion}

\subsection{The inverse Laplace transform}
Next, we evaluate the solutions $u_{\alpha \alpha'}(t)$ of the classical  equations of motion (\ref{sys}). Since Eq. (\ref{sys}) is linear, it can be solved by a Laplace transform. It is straightforward to evaluate the Laplace transform $\tilde{u}(z)$ of $u(t)$ as $A^{-1}(z)$, where $A(z)$ is the $2\times 2$ matrix with elements

\bea
A_{\alpha \alpha'}(z) =(z^{2}+\Omega_{\alpha}^{2}) \delta_{\alpha\alpha'}+ 2\widetilde{\gamma}_{\alpha\alpha'}(z),
\eea
where $\tilde{\gamma}_{\alpha\alpha'}(z)$ is the Laplace transform of the dissipation kernel. The Laplace transforms of $\gamma_0$ and $\gamma_r$ are

\bea
\tilde{\gamma}_{0}(z)&=&-\frac{\lambda^2}{16\pi ^2} \ln\left(1 +\frac{\Lambda^2 }{z^2} \right) \simeq -\frac{\lambda^2}{8\pi ^2} \ln\left( \frac{\Lambda }{z} \right)\\
\tilde{\gamma}_{r}(z)&=& -\frac{\lambda^2}{16\pi rz} [e^{-rz}\bar{E}i(rz)-e^{rz}Ei(-rz)],
\eea
where we simplified $\gamma_0(z)$ by assuming that the relevant values of $z$ satisfy $|z|<< \Lambda$; $\mbox{Ei}$ stands for the exponential integral function,  defined by \cite{Abramowitz}

\bea
Ei(z)=\gamma + \ln z +\sum_{1}^{\infty}\frac{z^{n}}{n!n}
\eea
where $\gamma$ is the Euler-Mascheroni constant and $\bar{E}i(z)=Ei(\bar{z})$.

It follows that

\bea
\tilde{u}(z) = \frac{1}{2} \left[\frac{1}{z^2 + \Omega^2 + 2 \tilde{\gamma}_{0}(z) +  2\tilde{\gamma}_{r}(z)}  \left(\begin{array}{cc} 1&1\\1&1\end{array}\right)  \right. \nonumber \\
\left. + \frac{1}{z^2 + \Omega^2 + 2c\tilde{\gamma}_{0}(z) -  2 \tilde{\gamma}_{r}(z)} \left(\begin{array}{cc} 1&-1\\-1&1\end{array}\right)
\right]. \label{eq_A}
\eea
Hence, $u(t)$ takes the form,
\bea
u(t) = \frac{1}{2} \left[ f_{+}(t) \left(\begin{array}{cc} 1&1\\1&1\end{array}\right) +f_{-}(t) \left(\begin{array}{cc} 1&-1\\-1&1\end{array}\right) \right], \label{uti}
\eea
in terms of   functions $f_{\pm}(t)$ that is defined by  the  Bromwich integrals
\bea
f_{\pm}(t) =\frac{1}{2\pi i} \int_{c-i\infty}^{c+i\infty} dz  \frac{e^{zt}}{z^2 + \Omega^2 + 2 \tilde{\gamma}_{0}(z) \pm 2 \tilde{\gamma}_{r}(z)}, \label{bromw}
\eea
where $c$ is a real constant larger than the real part of any pole in the integrand.

\begin{figure}[t!]
\centering
\begin{tikzpicture}[scale=0.5, every node/.style={scale=0.4}]
% Configurable parameters
\def\gap{0.2}
\def\bigradius{3}
\def\littleradius{0.5}

% Axes
\draw [thin,->] (-2.7*\bigradius, 0) -- (0.7*\bigradius,0);
\draw [thin,->] (0, -3.5*\bigradius) -- (0, 3.6*\bigradius);
%\draw [help lines] (-10,-10) grid (10,10);

%Labels
\node at (2.5,-0.2){$\text{Re}z$};
\node at (-0.5,11) {$\text{Im}z$};
%\node at (-1.6,2.1){$a$};
\node at (1.8,9.7){$c+iR$};
\node at (1.8,-9.7){$c-iR$};

\node at (1,-10){A};
\node at (1,10){B};
\node at (-7.9,0.6){C};
\node at (-0.3,0.6){D};
\node at (-0.3,-0.6){E};
\node at (-7.9,-0.6){F};

\node at (-4.2,8.5){$C_{R_A}$};
%\node at (-7.7,3.4){$C_{R_B}$};
\node at (-8,-3.4){$C_{R_B}$};
\node at (0.6,0.3){$C_g$};

 \draw[line width=1pt,yshift=135pt,decoration={markings,
  mark=at position 0.5 with {\arrow[line width=1pt]{>}}},postaction={decorate}]
 (1,5) arc (90.8:153.5:8.4cm);

  \draw[line width=1pt,decoration={ markings,
  mark=at position 0.5 with {\arrow[line width=1pt]{>}}},postaction={decorate}] (-6.4,5.1) arc(149:184:8.4cm);

   \draw[line width=1pt,decoration={ markings,
  mark=at position 0.5 with {\arrow[line width=1pt]{>}}},postaction={decorate}] (-7.6,0.2) -- (0,0.2);

\draw[line width=1pt,decoration={ markings,
  mark=at position 0.5 with {\arrow[line width=1pt]{>}}},postaction={decorate}] (0,0.2) arc(90:-90:0.2cm);

       \draw[line width=1pt,decoration={ markings,
  mark=at position 0.5 with {\arrow[line width=1pt]{<}}},postaction={decorate}] (-7.62,-0.2) -- (0,-0.2);

         \draw[line width=1pt,decoration={ markings,
  mark=at position 0.5 with {\arrow[line width=1pt]{>}}},postaction={decorate}] (-7.6,-0.2) arc(174.8:213.3:8.4cm);

         \draw[line width=1pt,decoration={ markings,
  mark=at position 0.5 with {\arrow[line width=1pt]{>}}},postaction={decorate}] (-6.3,-5.5) arc(210:270.3:8.4cm);

  \draw[decoration = {zigzag,segment length = 3mm, amplitude = 0.8 mm},decorate] (-7.58,0)--(0,0);

\draw[line width=1pt,yshift=-10pt,decoration={ markings,
  mark=at position 0.2 with {\arrow[line width=1pt]{>}},mark=at position 0.85 with {\arrow[line width=1.2pt]{>}}},postaction={decorate}] (1,-9.36)--(1,10.114);

  \draw [fill=purple,purple] (-4,0.8) circle [radius=0.12];
    \draw [fill=purple,purple] (-4,-0.8) circle [radius=0.12];
  \draw [fill=purple,purple] (-2.5,0.8) circle [radius=0.12];
  \draw [fill=purple,purple] (-2.5,-0.8) circle [radius=0.12];

  \node at (-4,1.3){$z_+^{(+)}$};
  \node at (-4,-1.3){$z_+^{(-)}$};
  \node at (-2.5,1.3){$z_-^{(+)}$};
  \node at (-2.5,-1.3){$z_-^{(-)}$};

 \end{tikzpicture}
   \caption{Bromwich contour, branch cut and poles related to Eq. (\ref{bromw}). Integration is along a straight line from $c-i\infty $ to $c+i\infty$, where c is a real constant larger than the real part of the poles of the integrand. The contour is closed by a semicircle of radius $R\rightarrow\infty$.}
\label{fig:inverse_lapl}
\end{figure}
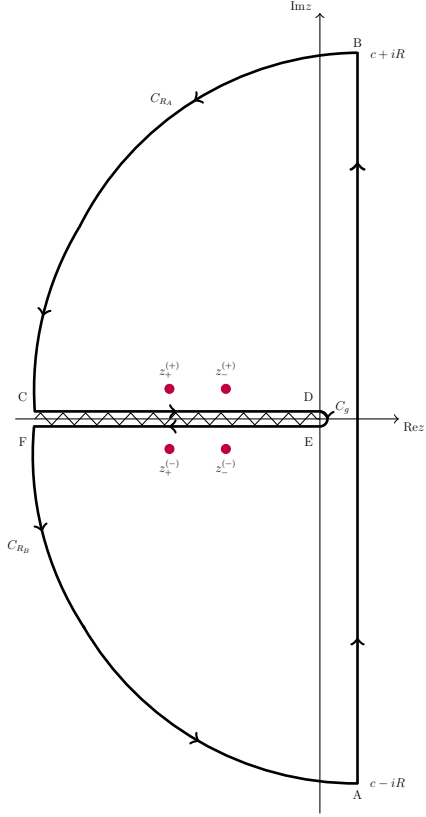

The integrand in Eq. (\ref{bromw}) has a branch cut at $z = 0$. For this reason, we consider  the integration  contour   of Fig. \ref{fig:inverse_lapl}--- that circles around the branch cut.
Using Cauchy's theorem, we find that the functions $f_{\pm}(t)$ consists of two parts,

\bea
f_{\pm}(t)= f^0_{\pm}(t) + I_{\pm}(t). \label{split}
\eea
The part $f^0_{\pm}(t)$ contains the contribution from the poles in the region enclosed by the contour, as in Fig. \ref{fig:inverse_lapl}---we will refer to it as the {\em pole term}. The part $I_{\pm}(t)$ includes the contribution from the negative imaginary axis; we refer to this term as the {\em branch-cut term}.

\subsection{The pole term}
 For sufficiently small $\lambda$, the poles can be identified perturbatively. To this end, we set $z_{+}^{\pm}= \pm i \Omega + \lambda^2 x$, and we solve the equation
\begin{eqnarray}
z^2 + \Omega^2 + 2 \tilde{\gamma}_{0}(z) \pm 2 \tilde{\gamma}_{r}(z) = 0 \label{poleq}
\end{eqnarray}
 to leading order in $\lambda^2$. We find that the poles associated to $f_+$ are at $z_{+}^{\pm} = \pm i \Omega +i \delta \Omega_+ - \Gamma_+$ and the poles associated to $f_-$ at $z_-^{\pm} = \pm i \Omega +i \delta\Omega_- - \Gamma_-$,
where
\begin{eqnarray}
\delta \Omega_{\pm} &=& -\frac{\lambda ^2}{8\pi ^{2}\Omega}\left( \ln\left(\frac{\Lambda }{\Omega }\right) \pm \frac{\cos(r\Omega )}{r\Omega}Si(r\Omega) \mp \frac{\sin(r\Omega )}{r\Omega}Ci(r\Omega)\right) \label{dom+-}\\
\Gamma_{\pm} &=& \Gamma_0\left(1 \pm \frac{\sin(r\Omega)}{r\Omega}\right) \label{gam+-} \\
\Gamma_0 &=& \frac{\lambda ^2}{16\pi \Omega }.
\end{eqnarray}
The constant $\Gamma_0$ is the decay rate of a single oscillator interacting with a scalar field.

Besides the two poles above, there exists a pole that is not accessible by perturbation theory. This solution corresponds to the regime $|z| << \Omega$. For example, consider the case that $r \rightarrow \infty$, so that the contribution of the $\tilde{\gamma}_r(z)$ term is negligible, Eq. (\ref{poleq}) has a root for $\mbox{Re} z \simeq \Lambda e^{- \frac{\pi \Omega}{2 \Gamma_0}}$. For finite $r$ the solution acquires an imaginary part. Since the real part of the root is positive, it
  leads to runaway solutions, i.e., it induces a term in $u(t)$ that blows up exponentially as $t \rightarrow \infty$. This term is   unphysical, because it is incompatible with the dissipative nature of the open system evolution. Its analogue appears in  the Abraham- Lorentz classical treatment of radiation reaction that leads to a third order equation for a particle's position \cite{Lor}. In fact, the exponentially runaway solution in this system  was first found by Planck \cite{Planck}. For the role of these solutions in QBM models of particle-field interaction, see, Ref. \cite{BaCa}.

These runaway solutions originate from the inadequacy of the particle-field  coupling to account for soft photons. In the present context, runaway solutions can be avoided by an infrared regularization. For example, we can regularize by assuming a finite mass $\mu$ for the scalar field. This is equivalent, to shifting the zero of $\gamma_0(z)$ by $\mu$, so that we redefine
\begin{eqnarray}
\gamma_0(z) = -\frac{\lambda^2}{16\pi ^2}  \ln\left(1 +\frac{\Lambda^2 }{(z + \mu)^2} \right).
\end{eqnarray}
For $\mu > \Lambda e^{- \frac{\pi \Omega}{2 \Gamma_0}}$, the third pole has a negative real part and does not lead to runaway solutions.    This regularization results to the integrand manifesting branch cuts at $z = -\mu \pm i \Lambda$, which have to be taken into account by an appropriate modification of the contour integral. In the weak coupling limit ($\Gamma_0/\Omega<< 1$), $\mu^{-1}$ is much larger  and $\Lambda^{-1}$ is much smaller than  physically relevant time-scales, so we can simply ignore the  contribution of this pole at physically relevant time scales. In contrast, for strong coupling, the runaway solutions cannot be  regularized away. The system of the two UdW detectors coupled with the scalar field is physically meaningful only in the weak coupling limit.

We conclude that  in the weak-coupling limit, the pole term is well approximated by, except at very early times ($t \sim O(\lambda^4)$).
\bea
f_{\pm}^{(0)}(t) = \frac{\sin \tilde{\Omega}_{\pm}t }{\tilde{\Omega}_{\pm}}e^{-\Gamma_{\pm}t}. \label{f+-0}
\eea

\subsection{ The branch-cut term}
To evaluate the integral along the negative near axis, we use the following identities.
\bea
\tilde{\gamma}_{0}(-s\pm  i \epsilon) &=& F(s) \mp i \frac{\lambda^2}{16 \pi} \\
\tilde{\gamma}_{r}(s \pm i \epsilon) &=& G(s)  \mp i \frac{\lambda^2}{16 \pi s r} \sinh(rs),
\eea
 for positive  $\epsilon \rightarrow 0$. The functions $F(s)$ and $G(s)$ are
 \bea
 F(s) &=& -\frac{\lambda ^{2}}{8\pi ^{2}}\ln\left(\frac{\Lambda }{s}\right) \\
 G(s) &=& -\frac{\lambda ^{2}}{8\pi ^{2}rs}\left[\cosh(rs)\mbox{Shi}(rs)-\sinh(rs)\mbox{Chi}(rs)\right]
 \eea
where $\mbox{Shi}$ is the hyperbolic sine integral function and $\mbox{Chi}$ the hyperbolic cosine integral function,  defined as

\bea
\mbox{Shi}(z)=\int_{0}^{t}\frac{\sinh(t)}{t}dt,
\hspace{1cm}
\mbox{Chi}(z)=\gamma +ln z + \int_{0}^{z}\frac{\cosh(t)-1}{t}dt.
\eea
Then,

\bea
I_{\pm}(t)=-\frac{\lambda ^{2}}{8\pi ^{2}}\int_{0}^{\infty} ds e^{-st}\frac{1\pm \frac{\sinh(rs)}{rs}}{(s^{2}+\Omega ^{2}+2F(s)+2G(s))^{2}+\left(\frac{\lambda ^{2}}{8\pi}\right)^{2}\left(1\pm\frac{\sinh(rs)}{rs}\right)^{2}}.\label{Invers_Lapl}
\eea
 The function $I_{\pm}(t)$ cannot be evaluated analytically. A good approximation that is valid for $t > r$ is to ignore the terms of order $\lambda^2$  in the denominator, so that
\bea
I_{\pm}(t)=-\frac{\lambda ^{2}}{8\pi ^{2}}\int_{0}^{\infty} ds e^{-st}\frac{1\pm \frac{\sinh(rs)}{rs}}{(s^{2}+\Omega ^{2} )^{2}} \label{it-}
\eea
For $t <r$, the approximation above does not hold, because dropping the terms of order $\lambda^2$ in the denominator renders the integral divergent.

For $\Omega t >> 1$, Eq. (\ref{it-}) becomes
\bea
I_{\pm}(t) = -\frac{\lambda ^{2}}{8\pi ^{2}\Omega^4} \left[ \frac{1}{t} \pm \frac{1}{r} \tanh^{-1}(r/t)\right].
\eea

\begin{figure}[t!]
\begin{center}
\subfloat[$\Omega I_+$ for $\Omega r=1$]{\includegraphics[scale=0.3]{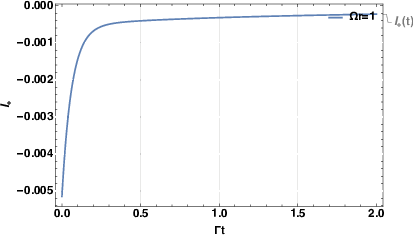}}\hspace{1cm}
\subfloat[$\Omega I_+$ for  $\Omega r=10$]{\includegraphics[scale=0.3]{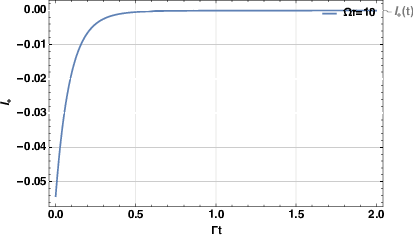}}\hspace{1cm}
\subfloat[$\Omega I_+$ for  $\Omega r=1000$]{\includegraphics[scale=0.3]{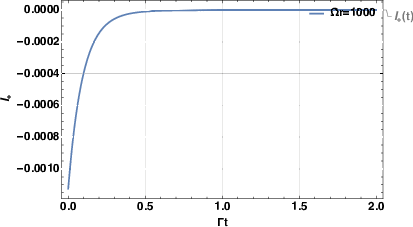}}
\hspace{1cm}
\subfloat[$\Omega I_-$ for $\Omega r=1$]{\includegraphics[scale=0.3]{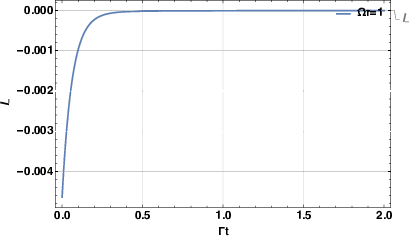}}\hspace{1cm}
\subfloat[$\Omega I_-$ for $\Omega r=10$]{\includegraphics[scale=0.3]{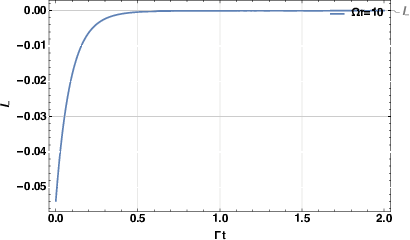}}\hspace{1cm}
\subfloat[$\Omega I_-$ for $\Omega r=1000$]{\includegraphics[scale=0.3]{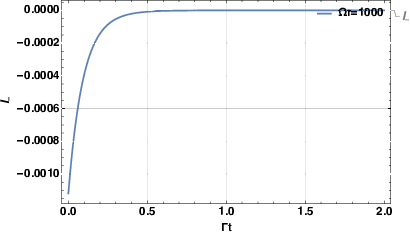}}
\hspace{1cm}
\end{center}
\caption{Evolution of $\Omega I_{\pm}$ as a function of $\Gamma_0 t$ for different values of $\Omega r $, where $\Gamma_0/\Omega=10^{-3}$.}
\label{Non_Mark}
\end{figure}

In Fig.(\ref{Non_Mark}) we  plot  $I_{\pm}$ as a function of $\Gamma_0 t$ for different values of $\Omega r$.
It is negative-valued and increases asymptotically to zero. It is unlike the pole term, in that it does not involve any oscillations.

\subsection{The Markov approximation}
Eq. (\ref{bromw}) is similar to the equation for the persistence amplitude of an unstable quantum state in the random phase approximation \cite{Ana19}. In fact, the two kernels $\tilde{\gamma}_0$ and $\tilde{\gamma}_r$ are similar  to the ones that appear in the evolution of a pair of atomic qubits interacting with the EM field \cite{ASH10}. The difference is that the dominant term contains a quadratic rather than a linear term with respect to $z$, reflecting that in a harmonic oscillator we consider both positive frequency and negative frequency solutions.

The split (\ref{split}) into a pole term and a branch-cut term is  generic whenever the kernels describing the effect of the environment contain branch-cuts. A common approximation in the study of unstable systems is the Wigner-Weisskopf approximation (WWA), in which (i) the branch-cut term is neglected, and (ii) the poles are calculated to leading-order in perturbation theory \cite{Ana19}. The WWA approximation leads to exponential decay. It coincides with van-Hove limit, namely, taking the limit $\lambda \rightarrow 0$, with $\lambda^2 t$ kept constant.  In the open quantum system context, the van Hove limit leads to  the second-order master equation that describes Markovian dynamics \cite{Davies}.

It is straightforward to evaluate the van Hove limit of Eq. (\ref{bromw}). A function of the form
\begin{eqnarray}
f(t) =\frac{1}{2\pi i}  \int_{c-i\infty}^{c+i\infty} dz  \frac{e^{zt}}{z^2 + \Omega^2 +  \lambda^2 a(z)},
\end{eqnarray}
for some kernel $ \lambda^2 a(z)$, can be written as
\begin{eqnarray}
f(t) = \frac{1}{2\pi i}  \int_{c-i\infty}^{c+i\infty} \frac{dz}{i\sqrt{\Omega^2 + 2 \lambda^2 a(z)}} \left[ \frac{1}{z- i \sqrt{\Omega^2 + 2\lambda^2 a(z)}} -   \frac{1}{z+ i \sqrt{\Omega^2 + 2\lambda^2 a(z)}}\right].
\end{eqnarray}
We set  $z = i\Omega + \lambda^2 x$ in the first term and $z = - i\Omega + \lambda^2 x$ in the second. Then, we take the limit $\lambda \rightarrow 0$, with $\lambda^2 t$ constant, to obtain
\begin{eqnarray}
f(t) = \frac{1}{\Omega} \left( e^{- i \Omega t - \frac{\lambda^2 a(i\Omega)}{\Omega}t} - e^{ i \Omega t - \frac{\lambda^2 a(-i\Omega)}{\Omega}t}\right),
\end{eqnarray}
 i.e., the pole term with a perturbative evaluation of the poles.

The van Hove limit essentially substitutes the classical equation of motion with non-local in time dissipation, with an equation that is local in time. Hence, it removes memory effects from the evolution equation. A local-in-time equation for dissipation is a necessary---but usually not a sufficient condition---for   Markovian dynamics. This can be seen in path integral derivations of the QBM master equation \cite{CaLe, HPZ}; Markovian behavior requires that the noise kernel also becomes local.

To summarize, the Markov approximation to the system under study presupposes the validity of the WWA approximation. Hence, the violation of the latter is a definite sign of the existence of non-Markovian dynamics.

\subsection{Non-Markovian dynamics}
The WWA, and consequently, the exponential decay law, cannot be valid at all times---see, the reviews \cite{decay1, decay2, Ana19}.  Exponential decay fails at very early times due to quantum Zeno dynamics. It also fails at very late times: the branch cut term typically falls off as an inverse power of $t$, and eventually becomes larger than the pole term that decays exponentially. However, the time scale for this decay is much larger than relaxation time. For example, in optical systems even for $\Gamma_0/\Omega$ as large as $ 10^{-3}$, the breakdown of the exponential decay takes place at $\Gamma_0 t \sim 30$, when less than $1:10^{26}$ of the initial systems remains in the excited state.

A  violation of the WWA is physically meaningful only if it takes place at time-scales compatible with the dissipation time, i.e., if it happens when $\Gamma_0 t$ is a small number. We will show that this takes place in the system studied here, when the detectors are separated by a large distance $r$.

Eq. (\ref{uti}) implies that $u_{11} = u_{22} = \frac{1}{2}(f_++f_-)$ and that $u_{12} = u_{21} = \frac{1}{2}(f_+- f_-)$. The terms $u_{11}$ and $u_{12}$ describe the dependence of the variables of one detector to the initial conditions of the second detector, while $u_{12}$ and $u_{21}$ essentially describe the correlations developed between the two detectors.

Eq. (\ref{dom+-}, \ref{gam+-}) imply that as $r\rightarrow \infty$, $\Gamma_+ = \Gamma_-$ and $\delta \Omega_+ = \delta \Omega_-$.
By Eq. (\ref{f+-0}), $f_+^{(0)}(t) = f_-^{(0)}(t)$ as $r \rightarrow \infty$, for all $t$. Hence, the pole part  of $u_{12}(t)$ vanishes for all $t$  as $r \rightarrow \infty$. In contrast, the branch-cut term remains finite. By continuity, for any given $t$ there is a finite distance $r$, at which the branch cut term dominates over the pole term, and hence, the WWA fails.

\begin{figure}[t!]
\begin{center}
\subfloat[]{\includegraphics[scale=0.3]{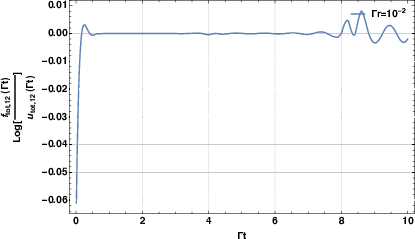}}
\hspace{1cm}
\subfloat[]{\includegraphics[scale=0.3]{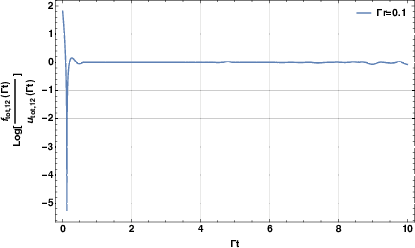}}
\hspace{1cm}
\subfloat[]{\includegraphics[scale=0.3]{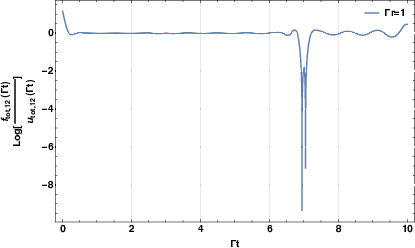}}
\hspace{1cm}
\subfloat[]{\includegraphics[scale=0.3]{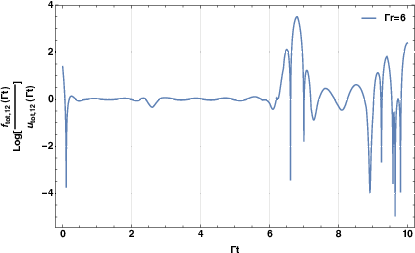}}
\hspace{1cm}
\subfloat[]{\includegraphics[scale=0.3]{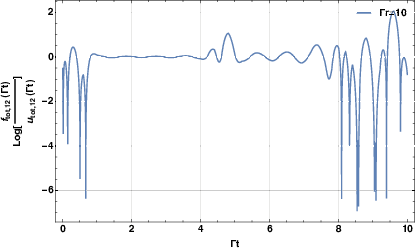}}
\hspace{1cm}
\subfloat[]{\includegraphics[scale=0.3]{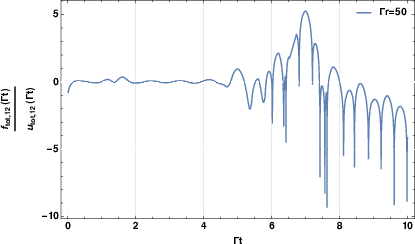}}
\hspace{1cm}
\subfloat[]{\includegraphics[scale=0.3]{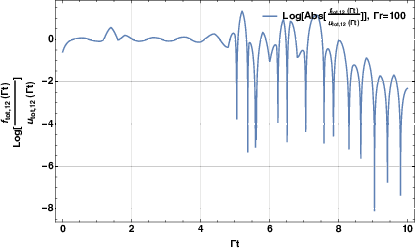}}
\hspace{1cm}
\subfloat[]{\includegraphics[scale=0.3]{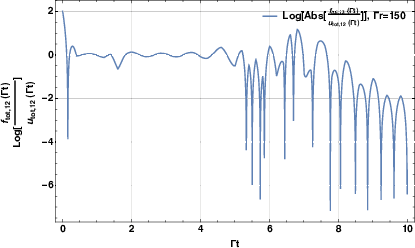}}
\hspace{1cm}
\subfloat[]{\includegraphics[scale=0.3]{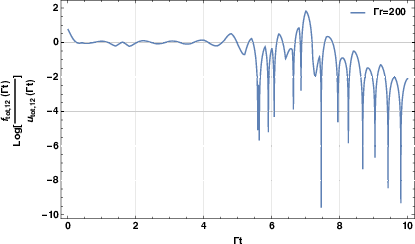}}
\end{center}
\caption{Evolution of the quantity $\frac{u_{12}^{(0)}}{u_{12}}$, where $u_{12}^{(0)}$ stands for the Markovian part of $u_{12}$, as a function of $\Gamma_0 t$ and for different values of $\Gamma r $. In this plot, $\Gamma_0/\Omega =10^{-3}$.}
\label{Mark.size_1}
\end{figure}

We have verified this behaviour numerically as can be seen in Fig. (\ref{Mark.size_1}). There, we present a semi logarithmic plot of  the  pole term of $u_{12}$ divided by the full $u_{12}$, as a function of time. We chose $\Gamma_0/\Omega = 10^{-3}$, i.e., we work well within the weak coupling regime. By construction, this ratio is very close to zero if the WWA holds, and it differs significantly from $0$ if the WWA fails. The plots show that the behavior of this function changes when  $r$ becomes of the order of $\Gamma_0^{-1}$. At this scale, we see significant violations of the WWA at the scale of $\Gamma_0 t \sim 1$ , and a complete breakdown as $\Gamma_0 t$ becomes about 5. Note that both violations and the breakdown of the WWA occur early, when a significant fraction of energy still remains in the system.

The WWA is well preserved for  $u_{11}$ and $u_{22}$  in the regime where it fails for $u_{12}$. Nonetheless, WWA also fails for  $u_{11}$ and $u_{22}$  at sufficiently large times. This is to be expected, because---as mentioned earlier--- the WWA is guaranteed to fail in the long time limit. What is rather unexpected, is that for sufficiently large $r$, the WWA breaks down at relatively early times also for $u_{11}$ and $u_{22}$. We found that for $\Gamma_0 r < 10$, the breakdown of the WWA occurs at $\Gamma_0 t \simeq 15$, i.e., at a time where a negligible amount of energy remains on the system. However, for $\Gamma_0 r > 50$, the WWA breaks down much earlier,  when $\Gamma_0 t \simeq 5$.

In all regimes that we have studied, the WWA breaks down at  the $u_{12}$ term both earlier and more strongly than it does at the $u_{11}$ and $u_{22}$ terms.  Therefore, the WWA fails primarily   for terms that describe the creation of correlation between distant detectors. For these terms, the  branch-cut contribution dominates. This result strongly suggests that the creation of correlations over large distances is a non-perturbative effect. It cannot be described correctly by perturbative approximation schemes, such as the von-Hove limit or the second-order master equation.

The conclusion above is unquestionable for the present model, because we have an exact solution, and consequently, full control over all approximation schemes. However, the open system evolution of the oscillator detector should not be significantly different from that of a $N$-level system coupled to a scalar field. For this reason, we expect that our conclusion is relevant to all systems with a similar Hamiltonian, in particular, to atoms coupled with a electromagnetic field. We have to go well beyond the second order master equation to describe the dynamics of entangled atoms, if these atoms are found at separations $r$ of order of $\Gamma_0^{-1}$.

The system   also exhibits non-Markovian behavior at the opposite regime $r \rightarrow 0$, as $\gamma_r \rightarrow \gamma_0$, and $f_-$ becomes simply $\frac{1}{\Omega} \sin \Omega t$. This  behavior has been extensively studied in multi-partite QBM models, see, for example, \cite{PaRo, LH08}. We will not be concerned with this regime here, because the limit $\Omega r << 1$ is not compatible with either the identification of the oscillators with atoms or with particle detectors.

 \section{Asymptotic states and generation of entanglement}
In this section, we show that the open system dynamics of the detectors lead to a unique asymptotic state. This state is correlated, and it is entangled for small separations.

\subsection{Asymptotic state}
In Sec. 2, we showed that the reduced density matrix propagator for this model is fully determined by the matrices $R(t)$ and $S(t)$. In Sec. 3, we evaluated $R(t)$ and showed its non-Markovian behaviour for $\Gamma_0 r \geq 1$. The matrix $S(t)$ is determined by Eqs. (\ref{sxx}---\ref{sxp}).

When evaluating the matrix elements $S_{ab}(t)$, we find that even for the non-diagonal elements the dominant contribution comes from the functions $u_{11}(t)$ and $u_{22}(t)$ and their derivatives. These functions are well described by the pole term except for very long times. Hence, we expect that the WWA is accurate for $S_{ab}(t)$. Numerically, we find that the difference between $S_{ab}$ calculated via the WWA and the exact expression is of the order of $\Gamma_0/\Omega << 1$.  If we substitute  solely the pole term for $u(t)$ in  Eqs. (\ref{sxx}---\ref{sxp}), integrations can be carried out analytically. They lead to an analytic expression for $S_{ab}(t)$ that is accurate to order $\Gamma_0/\Omega$.

The functions $u_{\alpha \alpha'}(t)$ vanish as $t \rightarrow \infty$, hence, so does the matrix $R_{ab}(t)$.  Eq. (\ref{equa130}) implies that as $t \rightarrow 0$, the Wigner function propagator becomes independent of $\xi_0$.
Numerical evaluation of $S_{ab}(t)$  shows that it asymptotes to a constant matrix for large $t$---we denote this matrix by $S(\infty)$.
Hence, asymptotically the system is described  by the Wigner function
\begin{eqnarray}
W_{\infty}(\xi) = \frac{1}{\pi \sqrt{\det S(\infty) }}\; \exp\left[-\frac{1}{2}S^{-1}_{ab}(\infty)\xi^{a}\xi^{b}\right],
\end{eqnarray}
By Eq. (\ref{equa2}), the correlation matrix at infinity $V_{ab}(\infty)$ coincides with $S_{ab}(\infty)$.

Interestingly, the matrix $S(\infty)$ involves correlations between the two detectors: the matrix elements $S_{X_1X_2}(\infty)$, $S_{P_1P_2}(\infty)$ and $S_{X_1P_2}(\infty)$ that describe such correlations are non-zero. To see this, we use the fact that   the dominant contribution to $S_{ab}(\infty)$ is well approximated by  the WWA. Substituting Eq. (\ref{f+-0}) into Eqs. (\ref{sxx}---\ref{sxp}), taking the limit $t \rightarrow \infty$, and keeping terms to leading order in
$\Gamma_0/\Omega$, we obtain
\begin{eqnarray}
S_{X_1X_1}(\infty) &=& S_{X_2X_2}(\infty) = \frac{\Gamma_0}{\Omega} \left[     \frac{1}{\Gamma_+} +   \frac{1}{\Gamma_-} - \frac{1}{2\Omega r} \left(\frac{\sin (\Omega_+r)}{\Gamma_+} -      \frac{\sin (\Omega_-r)}{\Gamma_-} \right)    \right] \label{sx1x1}
\\
S_{P_1P_1}(\infty) &=& S_{P_2P_2}(\infty) = \Gamma_0 \Omega \left[     \frac{1}{\Gamma_+} +   \frac{1}{\Gamma_-} -  \frac{1}{2\Omega r} \left(\frac{\sin (\Omega_+r)}{\Gamma_+} -    \frac{\sin (\Omega_-r)}{\Gamma_-} \right)    \right]\\
S_{X_1P_1}(\infty) &=& S_{X_2P_2}(\infty) = \frac{2\Gamma_0}{\Omega} \left(\frac{\delta \Omega}{\Omega} + \frac{ \sin (\Omega_+ r) - \sin (\Omega_-r)}{4\Omega r} \right)\\ \\
S_{X_1X_2}(\infty) &=& S_{X_2X_1}(\infty) = \frac{\Gamma_0}{\Omega}   \left[     \frac{1}{\Gamma_+} -   \frac{1}{\Gamma_-} - \frac{1}{2\Omega r} \left(\frac{\sin (\Omega_+r)}{\Gamma_+} +      \frac{\sin (\Omega_-r)}{\Gamma_-} \right)    \right]\\
S_{P_1P_2}(\infty) &=& S_{P_2P_1}(\infty) = \Gamma_0 \Omega \left[     \frac{1}{\Gamma_+} -   \frac{1}{\Gamma_-} -  \frac{1}{2\Omega r} \left(\frac{\sin (\Omega_+r)}{\Gamma_+} +       \frac{\sin (\Omega_-r)}{\Gamma_-} \right)    \right]\\
S_{X_1P_2}(\infty) &=& S_{X_2P_1}(\infty)  =\frac{ \Gamma_0}{\Omega} \left(-1 + \frac{ \sin (\Omega_+ r) + \sin (\Omega_-r)}{2\Omega r} \right) \label{sx1p2}
\end{eqnarray}
Remarkably, the correlation terms $S_{X_1 X_2}$ and $S_{P_1P_2}$ turn out to be of order $(\Gamma_0/\Omega)^0$, i.e., of the same order with the diagonal terms. However, unlike the diagonal terms, correlation terms are suppressed as $\Omega r $ becomes significantly larger than unity. For $\Omega r \simeq 20$ or smaller,  there is significant residual correlation between the detectors. This may appear surprising, but we note that the destruction of correlations at late times may be a common feature of either high-temperature baths, or systems of qubits, but it is {\em not} a generic property of open quantum systems. The existence of asymptotic correlations appears more intuitive  when viewing the oscillators as actual particle detectors. We would expect the detectors to develop correlations if they dominantly interact with particles with de Broglie wavelength of the order of their distance\footnote{There is no lower limit to $\Omega$ in our model---except for the infrared cut-off--- so the detectors could be correlated even if they are separated by macroscopically large distances. Of course, actual particle detectors are  macroscopic systems, and the variables $\hat{X}_{\alpha}$ are highly coarse-grained. The inclusion of additional degrees of freedom to the detector would introduce decoherence effects that would suppress such correlations beyond some length scale $L$.  }.

Next, we examine whether the    asymptotic state  is  entangled. To this end, we employ the Positive Partial Transpose (PPT) separability criterion of
 Peres and Horodecki \cite{Peres1996, Horodecki1997}.  In the present context, the PPT criterion  is applied to the correlation matrix $V$. A correlation matrix on $L^2({\pmb R}) \otimes L^2({\pmb R})$ is separable if it satisfies

\bea
V \geq -\frac{i}{2}\tilde{\Omega}, \quad \tilde{\Omega}=\Lambda \Omega \Lambda \label{simoncrit}
\eea
where $\Omega$ is the symplectic form on the four-dimensional phase space of two particles and $\Lambda$ is the matrix of the PPT operation $\Lambda = \mbox{diag}(1,1,1,-1)$ \cite{Simon}.

 In Fig. 4, we plot the minimal eigenvalue of $S(\infty)+ \frac{i}{2} \tilde{\Omega}$  as a function of $\Omega r$. A negative value of $\lambda_-$ indicates an entangled Gaussian state, a positive value a separable Gaussian state. We see that the asymptotic state is entangled for   $\Omega r \lessapprox 1.79$, and that the entanglement is stronger as $r \rightarrow 0$. The results are qualitatively  compatible with the analysis of Ref. \cite{Rez13} (that ignores backreaction) and the analysis of Ref. \cite{LH08} (that employs an expansion scheme). We note that  Eqs. (\ref{sx1x1}---\ref{sx1p2}) provide the exact asymptotic expression  of $S$ in the weak coupling limit.

 \begin{figure}[t!]
\begin{center}
 {\includegraphics[scale=0.3]{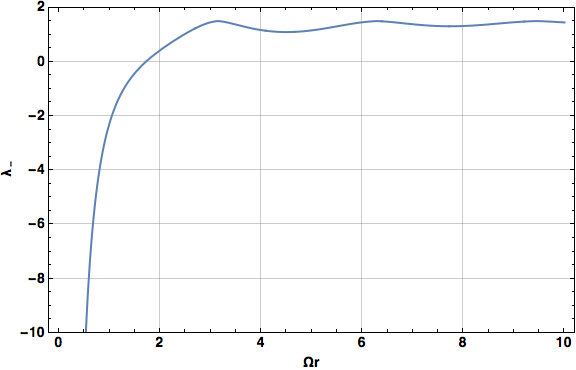}}
\end{center}
\caption{The minimal eigenvalue $\lambda_{-}$ of the matrix $S(\infty)+ \frac{i}{2} \tilde{\Omega}$ as a function of $\Omega r$.}
\end{figure}
\subsection{Entanglement generation}

Having established the asymptotic behaviour of the two-detector system, and identified the asymptotic behaviour of entanglement, we examine how entanglement is generated in time.  Again, we employ the separability criterion (\ref{simoncrit}). We consider an initial factorized state $|z\rangle \otimes |z'\rangle$ that is a product of coherent states. In Fig. 5, we plot the lowest eigenvalue of $V_t+ \frac{i}{2}\tilde{\Omega}$ as a function of $\Gamma_0 t$, where $V_t$ is given by Eq. (\ref{equa2}). As expected, entanglement is generated only at early times.

\begin{figure}[t!]
\begin{center}
\subfloat[$\Omega r=0.5$]{\includegraphics[scale=0.3]{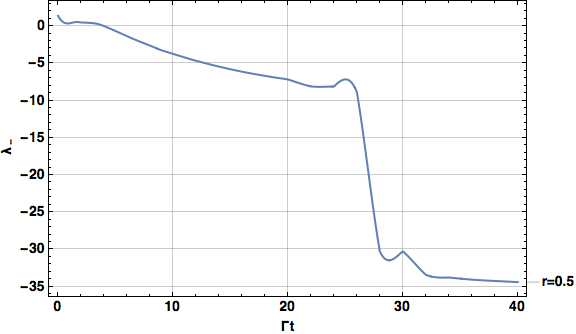}}\hspace{1cm}
\subfloat[$\Omega r=10$]{\includegraphics[scale=0.3]{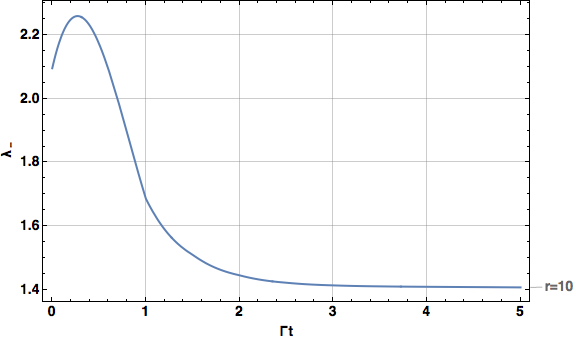}}\hspace{1cm}
\subfloat[$\Omega r=100$]{\includegraphics[scale=0.3]{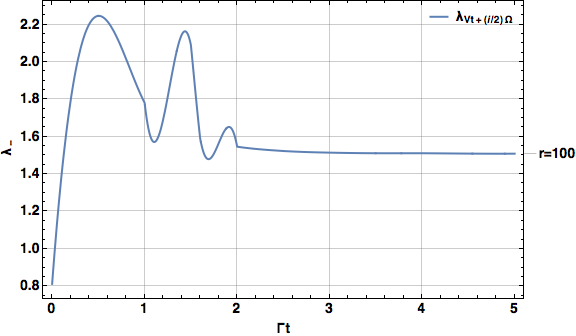}}\hspace{1cm}
\end{center}
\caption{The evolution of minimal eigenvalue $\lambda_{-}$ of $V_{t}+ \frac{i}{2}\tilde{\Omega}$ for initial factorized state $|z\rangle \otimes |z'\rangle$ and for different values of $\Omega r$. We see that entanglement is generated only for small $r$.}
\label{Harvest}
\end{figure}

 The choice of the initial state $|z\rangle \otimes |z'\rangle$ does not significantly affect the entanglement creation. Other factorized initial  states exhibit the same behaviour.

  For $z = z' = 0$, the initial state is  $|0, 0\rangle$, i.e., the ground state of the system of two oscillators. However, this state is not the lowest energy state for the full field-detector Hamiltonian. For this reason, the energy of the detector degrees of freedom momentarily increases as a result of the interaction with the environment, which would be paradoxical if $|0, 0\rangle$ were a true ground state.

 The state $|0, 0\rangle$ may be viewed as a ground state of the system if we can assume a set-up in which the field-detector coupling switches on at $t = 0$. As long as the switching on takes place at time-scales much smaller than $\Gamma^{-1}$, the solutions to the reduced dynamics derived here are applicable.

      In this context, the creation of entanglement from an initial vacuum state is referred to as {\em harvesting} of the QFT vacuum.   Most research on harvesting focuses on the evaluation of the effect in the lowest order of time-dependent perturbation theory. This is a good approximation as long as the interaction is switched on for a time interval much smaller than the relaxation time. For longer times, an open-quantum system treatment that takes back-reaction into account is essential, otherwise the effects of relaxation cannot be incorporated into the description.         For $\Omega r > 1.79$ that asymptotic state has only  classical correlations. This implies that studies of entanglement extraction that ignore backreaction may significantly overestimate the amount of harvested entanglement.

Finally, we note that  there is no significant generation of entanglement outside the light-cone  for static detectors.

 \section{The challenge of causality}
 An important motivation of this work is to understand how causality is implemented in the communication of separated localized quantum objects through a quantum field. The present model, being exactly solvable, provides an explicit  demonstration of Fermi's two atom problem,  in which the fundamental physical issues are not obscured by questions about the validity of  approximations.

 It is straightforward to verify that  the classical equations of motion (\ref{sys}) are not causal: $\hat{X}_2(t)$ depends on the value of $\hat{X}_1(0)$, even for  times $t < r$. This result is not surprising. Eq. (\ref{sys}) describes the interaction between the oscillators in terms of direct coupling in position---even if it is non-local in time---and it is well known that direct particle coupling cannot lead to causal dynamics in relativistic systems. The problem is that Eq. (\ref{sys}) describes the evolution of the expectation values of the observables $\hat{X}_{1,2}$, hence, its non-causal behavior seemingly implies superluminal signals.

Having an exactly solvable model allows us to demonstrate explicitly that this non-causal behavior is not an artefact of common approximations employed in such systems---for a treatment of causality violation in interactions between Unruh-DeWitt detectors, see \cite{Eduardo1, Eduardo2}. In particular, non-causality is not due to the choice of a factorizing initial condition, that was employed in   the derivation of the density matrix propagator. Such a condition  cannot hold exactly, because any preparation of the system cannot affect arbitrarily high energies of the field. Factorizability holds at most up to a cut-off energy scale. However, as mentioned in Sec. 2.1.1, existing models in the theory of quantum open systems strongly suggest such correlations are mostly significant at early times, and that their effects becomes negligible as correlations are established between system and environment due to dynamical interaction.

More importantly, we can derive an exact evolution equation for the expectation value  $\langle \hat{X}_r\rangle $  \cite{AKM10}
\begin{eqnarray}
\frac{d^2}{dt^2} \langle \hat{X}_{\alpha} (t)\rangle + \Omega_{\alpha}^2 \langle \hat{X}_{\alpha}(t)\rangle +2 \sum_{\alpha'} \int_0^t \gamma_{\alpha \alpha'} (t - s) \langle \hat{X}_{\alpha'}(s)\rangle = \sum_i \frac{c_{i \alpha}}{M_\alpha} \langle \hat{q}_{i}^{0}(t)\rangle,
\end{eqnarray}
where $\hat{q}_i$ is the field operator associated to the $i$-th mode, evolving according to the free equations of motion for the field. We can also choose  the  initial state to satisfy $\langle \hat{\phi}(x)\rangle = \langle \hat{\pi}(x)\rangle = 0$, where $\hat{\pi}(x)$ is the field conjugate momentum\footnote{This is a natural condition for a state that behaves like the field vacuum. In any case, the mean value of the field and its conjugate momentum can be shifted to any value by a unitary action of the Weyl group, that is generated by the field canonical algebra.}. This condition implies  that $\langle \hat{q}_{i}^{0}(t)\rangle = 0$, hence, $\langle \hat{X}_{\alpha}(t)\rangle$ satisfies  to Eq. (\ref{sys}). Mean values evolve non-causally, irrespective of the initial condition.

The situation  is analogous to that of Fermi's two-level atom that was mentioned in the introduction. In this sense, it is generic to all relativistic systems, when we attempt to describe their subsystems as completely localized in space. Hegerfeldt  proved with minimal assumptions that for any systems A and B, in disjoint regions, that interacting through a quantum field,  the excitation probability of B is nonzero immediately after $t= 0$ \cite{Heg}.  The present model exemplifies Hegerfeldt's theorem in an exactly solvable system.

Hence, this type of non-causality is not a feature of unphysical dynamics, for example, due to the limited validity of the field-particle coupling of this model. Another way to see this is the following.
  Field-particle couplings can be derived for the dynamics of an $N$-level atom coupled to the electromagnetic field \cite{W.M.2008}. The harmonic oscillators considered here can be viewed as atoms with equal spacing in the levels and $N \rightarrow \infty$.
   The starting point in such derivations is the full Quantum Electrodynamics. The  crucial condition that leads to couplings of the form (\ref{eq2}) is the {\em dipole approximation}. This asserts that the size of the localized systems is much smaller than the wavelength of the emitted radiation. Since the size of those systems defines the cut-off frequency $\Lambda$, the dipole approximation is expected to hold with an accuracy of the order of $\Omega/\Lambda$. Hence, corrections to the dipole approximation (and, hence to the field particle coupling)  are expected to increase with $\Omega$ and to be sensitive on the cut-off $\Lambda$. This is the case for the runaway solutions that are  regularized away---see, Sec. 3.2.
      In contrast, the non-causal behavior  that characterizes Eq. (\ref{sys}) is insensitive to $\Omega$ or to $\Lambda$.

For this reason, we believe that the problem of causality in detector-field interactions is fundamentally {\em kinematical} and not dynamical. This is supported by several theorems on the impossibility to define localization observables in relativistic quantum systems \cite{Schl71, Heg98, Mal96}.  Existing definitions of localized observables conflict   the requirement of relativistic causality. Observables that appear to be local and causal in classical theory or in on-relativistic quantum theory (e.g., a particle's position) fail to be so in relativistic quantum theory. In particular, this is the case for the quantities $\hat{X}_\alpha$ and $\hat{P}_\alpha$ that describe the degrees of freedom of the oscillator detectors in the present model. Once the interaction with the field is present, they cannot longer by viewed as localized observables pertaining to a single detector. Being non-local observables, their non-causal evolution is not problematic.

This also means that  a causal description  of relativistic transmission of information requires a consistent definition of localized observables. The Hilbert space of the total system is ${\cal H}_{tot} = {\cal H}_{d1}\otimes {\cal H}_{d2}\otimes{\cal H}_{field}$, where ${\cal H}_{d\alpha}$ is a Hilbert space associated to the $\alpha$ detector and ${\cal H}_{field}$ the field Hilbert space. An operator that corresponds to a measurement in the detector 1 should not be of the form $\hat{A} \otimes \hat{I} \otimes \hat{I}$, but rather it should be a non-factorized operator on ${\cal H}_{tot}$ that reduces to the factorizing form for $\lambda \rightarrow 0$. Rather heuristically, a local observable should include a contribution "virtual photons" in order to be compatible with causality \cite{Heg2}.

It is doubtful that self-adjoint operators that generalize $\hat{X}_\alpha$ and $\hat{P}_\alpha$ for the interacting system can be defined in a way that is compatible with causality. There are strong arguments that ideal measurements---i.e., measurements corresponding to self-adjoint operators---are incompatible with causality in QFT \cite{Sorkin}. These arguments are completely independent from the analysis Fermi's two-atom problem, they involve a QFT analysis of measurement. They strongly suggest that {\em all}   QFT measurements must be expressed in terms of Positive-Operator-Valued measures (POVMs).   One of us has proposed the use of time-extended observables for the description of particle localization \cite{AnSav19}. Time extended observables correspond to POVMs that partly depend upon the dynamics of the quantum system \cite{AnSav12}. Hence, a model with exactly solvable dynamics, such as the one analyzed here, is important for the explicit construction of such observables and for testing their causal behavior.

\bigskip

\noindent{\em Implications to entanglement generation.} We argued that operators of the form $\hat{A} \otimes \hat{I} \otimes \hat{I}$ cannot be viewed as corresponding to a local measurement of the first detector, and similarly for operators of the form $\hat{I} \otimes \hat{A} \otimes \hat{I}$ in relation to the second detector. However, the representation of   local measurements with operators of this form is a cornerstone of quantum information theory. In particular, it is a prerequisite for identifying entanglement as a quantum resource. Of course, this representation is based fundamentally on non-relativistic quantum physics. It does not directly apply to relativistic quantum systems, and it does not incorporate the severe restrictions on raised by QFT requirements of locality and causality.

Hence, there is no fundamental justification that the usual measures of entanglement between the detectors   define a genuine quantum resource. In particular, one cannot assert that these measures describe non-classical correlations between localized measurements. This point renders the the physical relevance of entanglement harvested by the vacuum, or  of entanglement generation outside the light-cone is questionable. While they may correspond to actual physical phenomena, their justification requires an analysis at a more fundamental level and requires the prior QFT definition of the localized observables that are being measured in actual experiments.

\section{Conclusions}
A new generation of quantum experiments  will allow us to test important issues at the foundations of QFT and of quantum information, pertaining to the principles of  causality and  locality and their relation to non-classical correlations like entanglement. Exactly solvable models, like the one analysed here, allow us to explore regimes that will be experimentally accessible, but they are not adequately described be described by usual approximation schemes, such as  the Markov  approximation or  the perturbative analysis of master equations.  Our conclusion that the  generation of correlations between subsystems at large separations is   a non-perturbative process is particularly important in relation to this context.

  We believe that the model  presented here   provides an important tool for addressing foundational issues in QFT, because it has a formal exact solution, and provides full mathematical control to all approximation schemes.  It
  may be used for constructing  of localized observables to address the Fermi problem,  for understanding causal propagation of signals/information in QFT, and for generalizing existing quantum information concepts to relativistic systems.

\section*{Acknowledgements}
TK acknowledges support from Greece and the European Union (European Social Fund- ESF) through the Operational Programme «Human Resources Development, Education and Lifelong Learning» in the context of the project “Scholarships programme for post-graduate studies - 2nd Study Cycle” (MIS-5003404), implemented by the State Scholarships Foundation (IKY). Both authors acknowledge support from the Research Committee of the University of Patras via the "K. Karatheodoris" program (Grant No. E611). We would also like to thank B. L. Hu, N. Savvidou, D. Moustos and K. Blekos for useful discussions and comments.

%==========BIBLIOGRAPHY=======================================================
%===========================================================================

 \end{document}